# Thermodynamics and kinetics of core-shell versus appendage co-precipitation morphologies: an example in the Fe-Cu-Mn-Ni-Si system

*Shipeng Shu[1], Peter B. Wells[2], Nathan Almirall[2], G. Robert Odette[2],*

*Dane D. Morgan[1],\**

[1] Department of Materials Science and Engineering, University of Wisconsin Madison

[2] Materials Department, University of California, Santa Barbara

\* Corresponding Author Email: ddmorgan@wisc.edu

**Abstract**

What determines precipitate morphologies in co-precipitating alloy systems? We focus on alloys of two precipitating phases, with the fast-precipitating phase acting as heterogeneous nucleation sites for a second phase manifesting slower kinetics. Kinetic lattice Monte Carlo simulations show that the interplay between interfacial and ordering energies, plus active diffusion paths, strongly affect the selection of core-shell verses appendage morphologies. We study a FeCuMnNiSi alloy using the combination of atom probe tomography and simulations, and show that the ordering energy reduction of the MnNiSi phase heterogeneously nucleated on a pre-existing copper-rich precipitate exceeds the energy penalty of a predominantly Fe/Cu interface, leading to initial appendage, rather than core-shell, formation. Diffusion of Mn, Ni and Si around and through the Cu core towards the ordered phase results in subsequent appendage growth. We further show that in cases with higher primary precipitate interface energies and/or suppressed ordering, the core-shell morphology is favored.

## 1. Introduction and background

Modern multiconstituent precipitation hardened alloys are designed to achieve the best combination of mechanical properties, thermal stability, corrosion resistance, etc. Controlling the detailed morphologies and structures of the precipitates is key to achieving optimized alloys. Very different precipitate morphologies have been observed in systems with two distinct precipitating phases [1]. These include core-shell structures [2, 3], second-phase appendages (side-by-side co-precipitate) [4-7], or simply two spatially separate populations of precipitates [8-10]. These microstructures are determined by the coupling of many complex thermodynamic and kinetic factors. For example, if the two precipitating phases wet each other, due to the reduction of total interfacial free energy, a co-precipitate morphology is favored. In many cases, one precipitate phase develops first, forming a template for the final co-precipitate structures. Further, by acting as heterogeneous nucleation sites, rapid formation of the template precipitates of one phase can promote the controlled nucleation of other phases that would otherwise not occur. The possibility of precisely controlling the initial template precipitate distributions holds the promise of opening a path to a wide range of new, high-performance microstructures. Co-precipitation generally leads to core-shell or appendage structures, which, however, often have dramatically different thermal stabilities [2, 6]. Thus, it is important to understand the underlying mechanisms of co-precipitation to optimize alloy designs. The focus of this work is specifically on the factors controlling core-shell verses appendage morphologies.

By coupling recent advances in near-atomic resolution characterization techniques in three dimensions, such as atom probe tomography (APT) [11, 12], with atomistic simulation tools such as kinetic lattice Monte Carlo (KLMC) [13], it is possible to study the precipitation pathways of multicomponent alloys in a quantitative manner. Here, we use these techniques to model the pathway of Cu-MnNiSi co-precipitation observed in irradiated (and thermally aged) Cu-bearing low-alloy steels [14-16]. Specifically, we focus on the mechanism of a transition of the co-precipitate from a Cu-core-MnNiSi-shell structure to a Cu-core-MnNiSi-appendage structure. The formation of appendage structures in this steel is representative of similar observations in other systems [4-7], but the underlying mechanisms driving appendage formation has not been elucidated.

Precipitation in Cu-bearing low-alloy steels has been intensely investigated for over 30 years, especially for the application to irradiation embrittlement of nuclear reactor pressure vessels steels. Since Cu is highly insoluble and diffuses rapidly, the initial precipitates are copper-rich, and often alloyed with Mn, Ni and Si. These precipitates form very rapidly, at concentrations above > 0.1 at.% around 300°C, due to radiation enhanced diffusion [17]. Previous thermodynamic models suggested that in addition to enriching the Cu core, solutes such as Mn, Ni and Si form a surrounding shell due to reduction of the high interfacial energy between Fe and Cu [18]. Such core-shell structures were subsequently widely reported in APT studies, both under irradiation and thermal aging conditions [16, 19-21]. More recently, APT showed that at longer times (or higher irradiation fluence) after Cu in matrix is depleted, and when more Mn, Ni and Si atoms come out of solution, the core-shell structure gives way to a Cu-core-MnNiSi-appendage co-precipitate morphology. However, the transition from core-shell to the appendage structure is counterintuitive, since it appears to create more high-energy Cu-Fe interface, compared to the normal concentric core-shell structure. In this article, we use the combination APT characterizations and KLMC simulations to study the Cu-MnNiSi two-phase precipitation, and demonstrate how the interplay of interfacial energy, ordering energy, and diffusion kinetics determines the detailed structure of the co-precipitate.

## 2. Methods

2.1 Materials

The alloy employed in the experiments is a Ni, Mn, Si, Cu-bearing steel with (in units of at.%) 0.25% Cu, 1.18% Ni, 1.08% Mn, 0.54% Si initially in solution. Note that the total Mn, Ni, Si content of 2.8 at.% is more than 10 times that of the Cu. The alloy was neutron irradiated to a fluence (E > 1 MeV) of $6.3\times10^{19}$ n cm$^{-2}$ at a flux of $1.0\times10^{14}$ at ≈ 300°C, and to a fluence of $1.4\times10^{20}$ n cm$^{-2}$ at a flux of $3.6\times10^{12}$ n cm$^{-2}$s$^{-1}$ at ≈ 290°C, in the Belgian Reactor 2 (BR2) and the US Advanced Test Reactor (ATR), respectively [14]. The precipitates observed in these alloy/irradiation conditions are fully representative of those in other Cu bearing steels with medium to high Ni contents for a wide range of irradiation and aging conditions [14, 22-24].

2.2 Atom probe tomography

APT samples were analyzed in a LEAP 3000X HR at the University of California, Santa Barbara. The APT measurements were performed in the voltage mode at 50K with a pulse fraction of 20% of the standing voltage, a 0.3-0.5% ion detection rate and a pulse repetition rate of 200 kHz [25]. The samples were fabricated using a dual-beam focused ion beam (FIB) system with the standard lift-out method as described in Ref. [26]. A final 2kV cleanup milling was carried out as the last step to reduce the Ga damage/contamination. SEM micrographs were used to calibrate the initial shape of the APT tips. Data reconstruction and analysis was performed using the Integrated Visualization and Analysis (IVAS) software. The atomic plane spacing in either (200) or (110) poles was used to scale the reconstructions in the z-direction. The precipitate analysis was performed using the cluster search tool in the IVAS software with order = 5 and $d_{max}$ = 0.5-0.6 nm. More details regarding the reconstruction method and parameters can be found in Refs. [14] and [17].

2.3 Kinetic lattice Monte Carlo simulations

The KLMC model was developed based on the model by Bellon and coworkers [27, 28], adding multinary alloy simulation capability and body-centered cubic (bcc) structural information. To simulate the precipitation process, the KLMC simulation was constructed with five alloying elements, Fe, Cu, Mn, Ni and Si, on rigid bcc simulation cells with sizes ranging from 64×64×64 to 128×128×128. The atomic bonds are modeled in the regular solution approximation, based on CALPHAD energies and fitting to post-irradiation annealing data, in terms of pairwise interaction energies $\varepsilon_{ab}$, $a, b$ = (Fe, Cu, Mn, Ni, Si). A single mobile vacancy is introduced into the system. The evolution of the precipitate structure is then a direct result of the vacancy exchanging its position with solute atoms, at an exchange frequency, $k_a^v$, between the vacancy and an atom on a nearest-neighbor lattice site. The exchange frequency is calculated using transition-state theory as $k_a^v = \nu_a \exp(-E_a^v/kT)$, where $\nu_a$ is the attempt frequency ($6 \times 10^{12} s^{-1}$) and $E_a^v$ is the activation energy of the exchange. The simulation time is incremented using a standard residence-time algorithm [29]. More details of the KLMC model and parameterization of the atomic interaction and activation energies are described in Ref. [30].

In the KLMC simulations, the frequency of the thermal jumps was determined using standard rate theory. The activation energy was calculated using the final-to-initial-system-energy (FISE) model [31] as

$$E_a^v = \frac{E_f - E_i}{2} + E_0^a,$$

where $E_f - E_i$ is the energy difference between the energies of the final and initial configurations for a certain vacancy-atom exchange, $E_0^a$ is the reference activation energy, assumed to be dependent only on the chemical species of the migrating atom, and was taken from Ref. [32].

$E_i$ and $E_f$ were calculated by evaluating the interaction energies between the vacancy/atom and their nearest neighbors. Homo-atomic pair interactions $\varepsilon_{aa}$ (i.e., interactions between atoms of the same chemistry) were related to cohesive energies through $E_{coh}^a = \frac{Z}{2}\varepsilon_{aa}$, where $Z$ is the nearest-neighbor site coordination number ($Z = 8$ for bcc structure). Hetero-atomic interactions $\varepsilon_{ab}$ (i.e., interactions between atoms with different chemistry) were defined through the ordering energy as $\omega_{ab} = 2\varepsilon_{ab} - \varepsilon_{aa} - \varepsilon_{bb}$. The value of $\omega_{ab}$ determines the shape of the binary $a$-$b$ phase diagram. Effective atom-defect pair interactions were used to reproduce the values of vacancy formation energies, defined as $E_{av}^f = Z\varepsilon_{av} - \frac{Z}{2}\varepsilon_{aa}$ [33].

The homo-atomic pair interactions $\varepsilon_{aa}$ were determined from measured cohesive energies [34] and DFT calculated values [35, 36] for bcc phase of the pure element. The hetero-atomic pair interactions $\varepsilon_{ab}$ were obtained from molar excess free energies ($G_{ab}^m$), calculated by the CALPHAD method [37]. Specifically, assuming a regular solution model, one can write

$$G_{ab}^m = x_a G_0^a + x_b G_0^b + RTx_a \ln x_a + RTx_b \ln x_b + x_a x_b \Omega_{ab},$$

where $\Omega_{ab} = \frac{Z}{2} N_A \omega_{ab}$. $\Omega_{ab}$ can be fit from the CALPHAD model for $G_{ab}^m$ and connects the CALPHAD outputs to the KMC inputs. Details on the process of determining $\Omega_{ab}$ (a,b = Fe, Mn, Ni and Si) can be found in Ref. [38]. Cu related interactions were obtained from Ref. [18]. The complete energetics used in this study are listed in Table 1, Table 2 and Table 3. Note that, due to the much lower Cu vacancy formation energy compared to that of the Fe matrix, it is expected that the vacancy would spend most of the time jumping inside the Cu precipitate. As a result, to achieve significant diffusion outside the Cu precipitate, very long simulations are required.

Table 1. Cohesive energy of elements used in KLMC simulations (all values are eV). [39]

| Element | Fe | Cu | Mn | Ni | Si |
|---|---|---|---|---|---|
| $E_{coh}^a$ | -4.28 | -3.49 | -2.92 | -4.34 | -4.03 |

Table 2. Vacancy formation energy in different elements (all values are eV). [35, 36, 39, 40] Note that the Si vacancy formation energy has been adjusted so that the binding energy matches the ab initio calculations. [35]

| Element | Fe | Cu | Mn | Ni | Si |
|---|---|---|---|---|---|
| $E_{av}^f$ | 1.60 | 0.90 | 1.40 | 1.48 | -0.21 |

Table 3 The interaction parameters used in the simulation (all values are eV). [38]

| $\Omega_{AB}$ | Cu | Mn | Ni | Si |
|---|---|---|---|---|
| Fe | 0.458 | 0.094 | 0.007 | -1.542 |
| Cu |  | 0.090 | 0.106 | -0.344 |
| Mn |  |  | -0.460 | -1.094 |
| Ni |  |  |  | -2.030 |

Certain irradiation effects are not included in the model. In particular, forced chemical (ballistic) mixing is not considered. Ballistic mixing is expected to be negligible at the relatively low ATR dose rate [41], but some ballistic mixing may occur at the higher BR2 dose rate. Nevertheless, since core-shell structures are commonly observed at lower flux, the BR2 results are generally representative of lower fluence irradiation conditions. Heterogeneous nucleation within cascades [42] is also not modeled, since it is expected that this would have little influence on well-developed appendage morphologies, and since the appendages form on the previously precipitated copper cores. In general there is no reason to believe that ballistic mixing or heterogeneous nucleation within cascades is relevant for the appendage structure formation as similar structures are also observed in thermally aged alloys [20]. Note that the vanishingly small concentration of defects (albeit much larger than equilibrium quantities) has no effect on the basic alloy chemical thermodynamics. Thus, the difference between co-precipitation and appendage formation thermally and under irradiation is simply a matter of absolute time scales. Our model treats the relative rates of various local solute transport processes, focusing on the thermodynamic and kinetic factors that result in the appendage structure itself, and there is no effort or need to simulate absolute time scales.

## 3. Results and discussion

3.1 Morphological characteristics of Cu-MnNiSi co-precipitate

Examples of atom maps from the APT reconstructions are shown in Figure 1 for the (a) lower fluence BR2 and (d) higher fluence ATR conditions, respectively. Note that the Mn, Ni and Si atoms are partially transparent to better show the precipitate morphology. The number densities ($N$), mean radii ($<r>$), mole fractions ($f$) and overall composition of the co-precipitates are summarized in Table 4. The precipitates in the lower fluence, higher flux BR2 condition are much more numerous and smaller than in the ATR case. The difference in the $N$ and $<r>$ is likely caused by both the lower fluence and the higher flux of BR2 condition. The higher flux leads to reduced effective fluence due to enhanced recombination and subsequently lower radiation enhanced diffusion, and may also lead to some ballistic mixing [28, 30, 43]. These results are consistent with previous experiments that have shown higher neutron flux results in smaller, more numerous precipitates at a given fluence [44]. It should also be noted that higher flux decreases the effective fluence relative to lower flux ATR irradiations [43]. In the case of the BR2 irradiation, the effective fluence is reduced by a factor of ≈ 2 with respect to the ATR condition. Thus, the precipitates in BR2 condition are in an earlier stage of development than those from ATR. However, since the Cu cores of the precipitates are well-developed in both conditions, they can be used to assess how MnNiSi phase evolves on the Cu cores as a function of the irradiation fluence.

The precipitates are predominantly Cu, enriched with some Mn, Ni and Si in the lower-fluence BR2 condition, with likely a core-shell structure. A magnified precipitate is shown in Figure 1b. Closer examination of the precipitates shows that the MnNiSi shell does not fully cover the Cu precipitate. Considering the APT detection efficiency of 37% one cannot meaningfully distinguish a significant fraction partial monolayer from a full monolayer, but the results are fully consistent with the LKMC simulations (see Sec. 3.2). A 1-D elemental concentration profile through the precipitate is shown in Figure 1c. This core-shell structure is consistent with many previous studies [18, 20, 45]. Note that the profile suggests a significant amount of Fe at the center of the precipitate. However, as thoroughly discussed in the literature, this high Fe concentration is mainly caused by trajectory aberration in the APT measurement [46-51]. It was recently shown by Shu et al. that the true Fe concentration in the Cu precipitate at ~ 300°C is less than a few percent [17]. At higher fluence, as shown in Figure 1d, the precipitate size increases by ≈ 60%. Due to rapid Cu precipitation being almost complete in both the high and low fluence conditions, the size of the Cu core is approximately the same in both cases. The further precipitate growth is almost entirely due to the addition of Mn, Ni and Si atoms. Most notably, the Mn, Ni and Si do not simply thicken the shell, but rather form an almost pure MnNiSi appendage attached to one side of the core, as shown in Figure 1e. The remainder of the Cu-Fe core interface is still loosely coated by Mn, Ni and Si atoms. A 1-D elemental concentration profile across the precipitate is again shown in Figure 1f, demonstrating that the center of mass of the Cu and the MnNiSi appendage are separated by a significant distance. The origin of this precipitate morphology transformation has not previously been reported, and is the focus of modeling in the following sections.

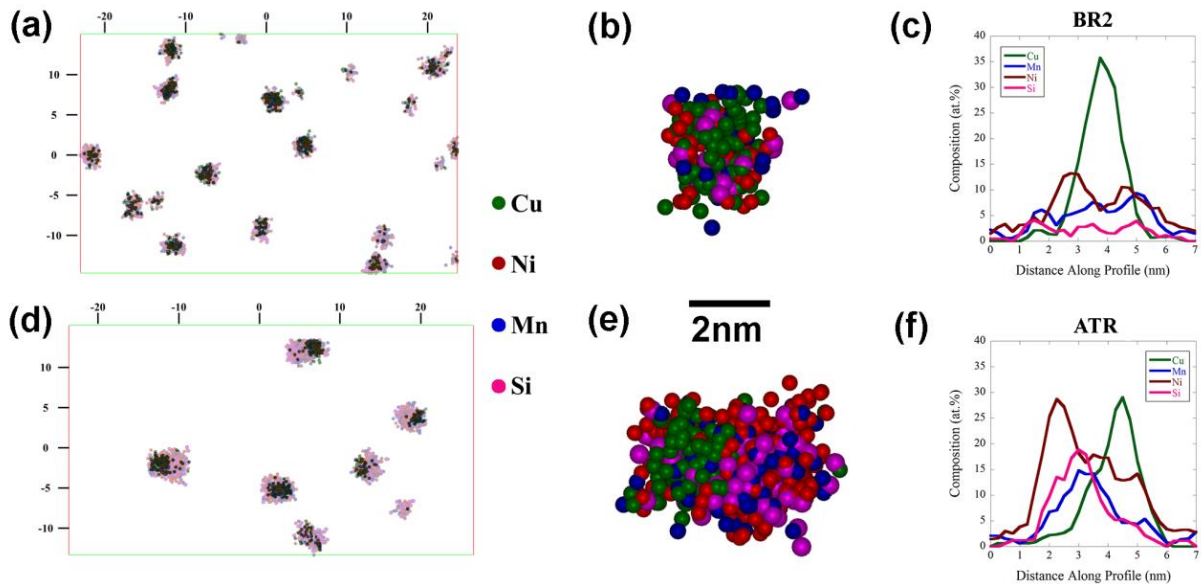

Figure 1 (a) atom maps from the BR2 condition (b) a magnified precipitate, BR2 condition (c) typical 1-D line profile of a precipitate of BR2 condition (d) atom maps from the ATR irradiation (e) a magnified precipitate, ATR condition (f) typical 1-D line profile of a precipitate of ATR condition. In the atom maps, Cu atoms are shown in solid green; Ni (red), Mn (blue) and Si (magenta) atoms are shown as partially transparent to more clearly see the precipitate morphology. (For interpretation of the references to color, the reader is referred to the web version of this article.)

Table 4 Number density ($N$), average size($<r>$) and volume fraction ($f$) of Cu-MnNiSi precipitates in samples irradiated in BR2 and ATR conditions.

|     | $N$ (m$^{-3}$) | $<r>$ (nm) | $f$ (at.%) | *Average Composition* |
| --- | --- | --- | --- | --- |
| BR2 | $9.5 \times 10^{23}$ | 1.1 | 0.57% | $Cu_{28}Ni_{36}Mn_{23}Si_{13}$ |
| ATR | $5.7 \times 10^{23}$ | 1.6 | 0.94% | $Cu_{17}Ni_{43}Mn_{21}Si_{19}$ |

3.2 KLMC simulation of the nucleation and growth of the Cu-MnNiSi precipitates

To study the evolution of the precipitate morphology, we allowed the simulated system to evolve at 300˚C, starting from a random solid solution with a composition measured by APT. Figure 2a is a representative snapshot of the multitude of configurations produced by the KLMC simulation. Remarkably, the KLMC simulations show essentially the same co-precipitate morphologies that are observed experimentally. Specifically, the Cu cores and MnNiSi phase form co-precipitates, with Cu cores on opposite sides of each other. The magnified precipitate in Figure 2b shows the magnified co-precipitate structure, with a cylindrical IVAS region of interest (ROI) passing through. Note that the copper-rich core maintains a loose coating layer of Mn, Ni and Si on the side away from the large MnNiSi appendage, again similar to the experimental observation. Figure 2c shows a 1-D composition profile along the z-axis of the ROI defined in Figure 2b. The asymmetric peaks representing the copper-rich precipitate and the MnNiSi phase can be clearly seen.

The MnNiSi phase forms an ordered B2 structure in the simulation, where Ni and Mn occupy the two sublattices, with Si randomly replacing some Mn atoms. The resulting Mn-Si trade-off is consistent with experimental observations that the number of Mn plus Si is approximately equal to the number of Ni atoms [14, 52]. Note the bcc structure imposed by the rigid KLMC lattice differs from the face-centered cubic based G-phase [53] precipitates that have been indexed in x-ray diffraction experiments [54]. However, as reviewed by Millán et al. [55], a B2 ordered structure can be favored for nano-scale Mn-Ni precipitates, which is expected to be fully coherent with both the bcc transition phase copper-rich precipitate and the Fe matrix. When the MnNiSi precipitates grow to larger sizes, they are likely to transform to the ordered intermetallic G-phase [54]. Although our rigid lattice simulation cannot reproduce this structural transformation, it does capture the ordering in the MnNiSi phase. Specifically, ordered $Mn_{0.5}Ni_{0.5}$ is more stable than the corresponding disordered alloy by 0.11 eV/atom. Si has a modest impact on this ordering energy, as randomly substituting Si for Mn in the ordered phase changes the energy by only $0.12 \times c_{Si}$ eV/atom for dilute Si (here $c_{Si}$ is the concentration of Si in the whole precipitate, e.g., $c_{Si} = 0.5$ for $Mn_{0.5}Si_{0.5}$).

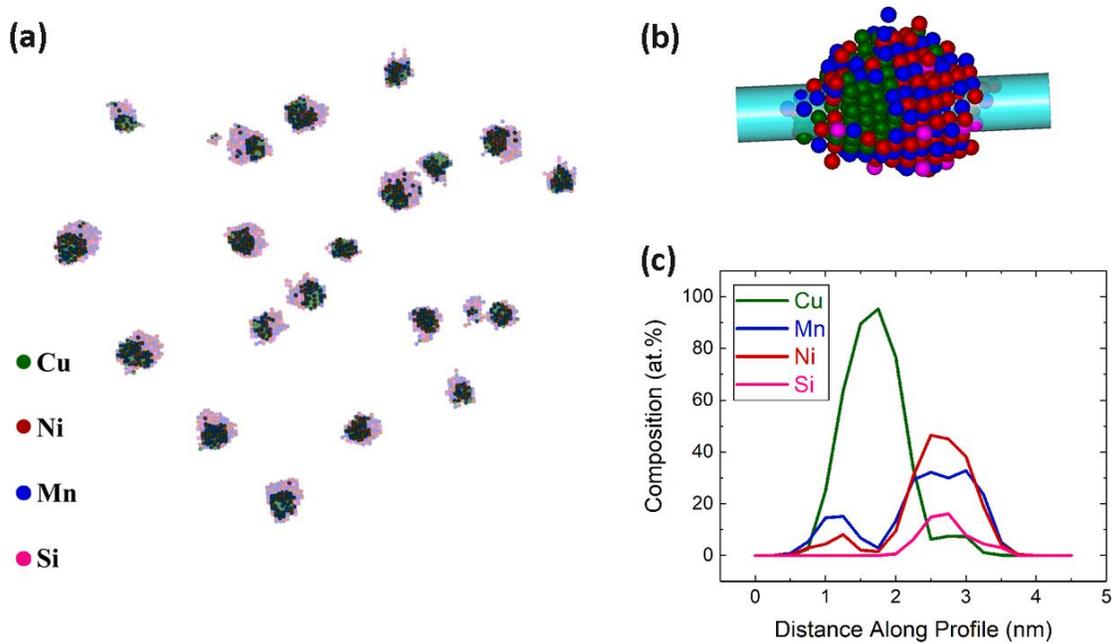

Figure 2 (a) a snapshot of the KLMC simulation, showing the microstructure of the simulated alloy, solute atoms in solution are not shown. Cu atoms are shown in solid green; Ni (red), Mn (blue) and Si (magenta) atoms are shown as partially transparent to more clearly see the precipitate morphology. (b) a magnified Cu-MnNiSi precipitate, with a cylindrical region of interest (ROI) along which a 1-D composition profile will be calculated (c) a 1-D composition profile along the z-axis of the ROI. (Color online)

KLMC simulations can track a particular precipitate and identify its evolution in the nucleation, growth and coarsening stages. Figure 3 is a sequence of snapshots of the evolution of a typical Cu-MnNiSi co-precipitate. At the early stage of formation, in Figure 3a, the precipitate has a copper-rich core, with a loose Mn, Ni and Si coating. The core-shell coating morphology is due to the resulting net decrease in Cu-MnNiSi plus Fe-MnNiSi versus the Cu-Fe interfacial energies [56]. At this stage, the small copper-rich precipitate is very mobile as a result of the very low vacancy formation energy in Cu [36]. At later stages of growth, as shown in Figure 3b, a MnNiSi appendage preferentially nucleates on one side of the previously formed copper-rich precipitate. As shown in Figure 3c-d, the MnNiSi precipitate appendage continues to slowly grow by incorporating Mn, Ni and Si atoms, forming an ordered phase attached to the copper-rich core. Note, the mobility of the copper core with the appendage is lower compared to the mobility at the initial stage of precipitation, likely due to a pinning effect imposed by the MnNiSi appendage. This initial high Cu cluster mobility followed by an onset of pinning may play an essential role in building quantitative models of Cu precipitate evolution in Fe-based alloys. The slower but much longer continued growth of the MnNiSi precipitates relative to Cu precipitates is due to the combination of factors: (a) the effective chemical potential of the Mn-Ni-Si solutes is much lower than that for Cu, thus lowering the corresponding flux to the precipitates; and, (b) there is more than 10 times more Mn + Ni + Si atoms initially in solution than Cu, most of which can accumulate in the MnNiSi co-precipitates over time. The simulation evolves to reproduce a MnNiSi appendage structure that is nearly indistinguishable from that observed in the APT experiments. Note that in the simulations, up to $6 \times 10^{12}$ jumps were employed for each run, thus representing $\approx 3 \times 10^6$ jumps per atom. We did not rescale the KMC time to real time, as it is difficult to accurately assess the radiation enhanced diffusion in the experiments. However, the Hamiltonian gave time evolution that agreed well with experiments in simulations of annealing [38].

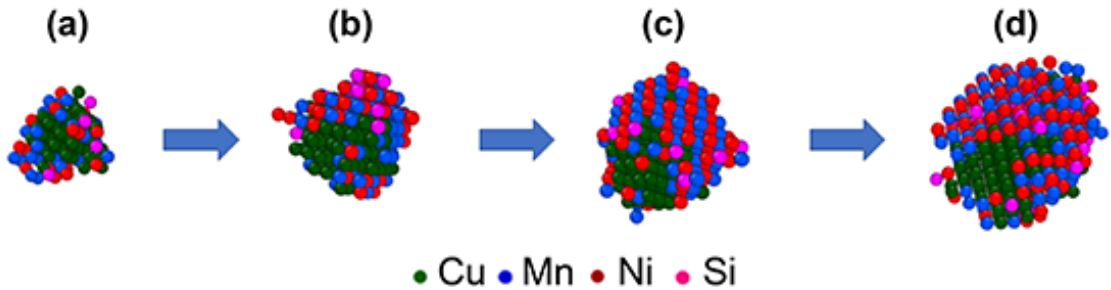

Figure 3 The evolution of the Cu-MnNiSi precipitate simulated by KLMC. (a) Formation of a Cu cluster coated by a layer of MnNiSi (b) nucleation of MnNiSi ordered phase on the Cu cluster (c) and (d) further growth of the MnNiSi ordered phase, note that the Cu cluster is always on one side of the precipitate. (color online)

To further demonstrate the path of asymmetric growth of the MnNiSi appendage, we measure the distance between the center of mass (COM) of the copper-rich core and that of the whole co-precipitate. Figure 4 shows that the distance between the COMs grows in proportion to the radius of the co-precipitate. These results clearly demonstrate that the COMs of the copper-rich core and whole co-precipitate move further and further apart, which would be expected for an appendage growth when copper-rich core is always on one side.

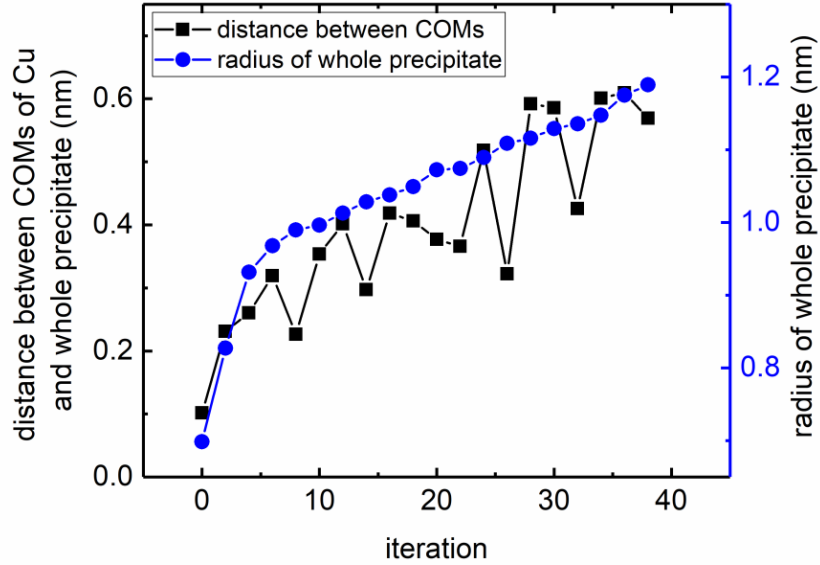

Figure 4 The distance between COMs as the precipitate grows, for the largest precipitate in the system. The radius of the whole precipitate is also plotted to show that the distance between COMs is directly correlated to size of the Cu-MnNiSi co-precipitate. In this specific simulation, one iteration corresponds to $\approx 1.6 \times 10^5$ jumps per atom. (color online)

3.3 Thermodynamics, kinetics and the mechanism of the structural transition

For a variety of reasons, it is known that the interfacial energy $\gamma_{Fe\text{-}Cu}$ is significantly higher than $\gamma_{Fe\text{-}MnNiSi}$. For example, MnNiSi precipitates would never thermally nucleate if $\gamma_{Fe\text{-}MnNiSi}$ was close to $\gamma_{Fe\text{-}Cu}$ [42, 57]. As a result of the very large supersaturation of even small amounts of Cu coupled with the decrease in the interface energy due to initial MnNiSi segregation, a high density of small copper-rich precipitates rapidly forms. Homogeneous nucleation of MnNiSi phase is insignificant except at very high Ni levels [14]. Instead, Mn, Ni and Si atoms slowly diffuse to the existing copper-rich precipitates. From Figure 1c it appears that the shell extends about 0.5 nm on each side of the 2 nm Cu core. Note that, considering the APT trajectory aberrations, this apparent thickness should be viewed as an upper limit [46]. In this limiting case, there would be more than 3 times as many Mn + Ni + Si than Cu atoms, although the shell thickness is only 25% of the diameter. As the layer formed near the interface is just a few atomic layers thick at most, it is likely that the Mn, Ni and Si atoms initially have no long-range order in the thin shell, consistent with the KLMC simulations. Considering a disordered MnNiSi layer, the X/MnNiSi (X = Fe, Cu) and Fe/Cu interfacial energies can be readily calculated from the KLMC Hamiltonian within the framework of the nearest-neighbor broken-bond (NNBB) model [58], and are listed in Table 5. According to the calculated interfacial energies, faceted precipitates are expected to form, with <110> facets dominating, which are indeed observed in the LKMC simulations.

Table 5 Interfacial energies for specific crystallographic orientations, estimated using the Nearest-Neighbor Broken-Bond model [58].

| Interfacial energies for specific crystallographic orientations (mJ/m$^2$) | Cu-Fe | Cu-MnNiSi | Fe-MnNiSi |
|---|---|---|---|
| [100] | 449 | 107 | 115 |
| [110] | 317 | 76 | 81 |
| [111] | 518 | 124 | 133 |

From classical wetting theory [59], the sign and magnitude of the spreading parameter $S = \gamma_{Cu-Fe} - (\gamma_{Fe-MnNiSi} + \gamma_{Cu-MnNiSi})$ determines how the MnNiSi phase wets the Cu core (total wetting, partial wetting, or complete drying for S > 0, -1 < S < 0, and S < -1, respectively). The interfacial energies listed in Table 5 give positive $S$ for all the crystallographic orientations, suggesting a total wetting state with contact angle being zero, which naturally would lead to the formation of a core-shell structure. Indeed, a thin, disordered, and dense MnNiSi layer on the Cu/Fe interface reduces the total interfacial energy by ~50%. A similar, but loose coating layer is actually observed in the lower-fluence BR2 experiment, as shown in Figure 1a-c. However, the shell does not thicken uniformly, but rather forms a small, ordered MnNiSi nucleus, located on one side of the Cu cluster. Interestingly, during further growth, the ordered phase never fully envelops the Cu core, which is kept on the edge of the whole precipitate, exposing the partially coated Cu-Fe interface. Next, we show that the formation of the appendage (and the suppression of the concentric core-shell structure) is a direct result of the ordering of the initial MnNiSi nucleus followed by the specific diffusion path of the Mn, Ni and Si atoms.

Since the original MnNiSi coating layer on Cu core is taken to be disordered (and thus provides no nucleation template for ordering), uniform nucleation and growth of an ordered thick MnNiSi shell is unlikely, as uniform nucleation requires the ordering of the shell be "in phase" on the whole surface to avoid the creation of anti-phase boundaries. Thus, it is more likely for an ordered nucleus to form locally on one facet of the Cu core (preferentially <110> facets, which are expected to be the dominating facets, according to the Wulff constructions of the crystals). In this situation, there are three distinct environments for the Mn, Ni, Si solute atoms: (1) dissolved in the matrix as monomers; (2) in the loose, disordered coating layer (which still lowers the interface energy by some extent); and (3) in the ordered MnNiSi nucleus. The solute atoms dissolved in the matrix and ordered phase have the highest and lowest energies, respectively. Thus, the latter would be expected to grow. However, if the Mn, Ni and Si solute atoms completely leave the disordered coating layer, a high-energy Cu-Fe interface is exposed, thus increasing the total free energy of the system. On the other hand, solute atoms leaving the disordered coating layer to join the ordered MnNiSi nucleus lower the total free energy of the system if the coating solutes are simply replaced.

It can be shown (see supplementary information) that there is a significant free energy penalty if all the Mn, Ni, Si atoms in the coating layer join the ordered nucleus, completely exposing the Cu/Fe interface. Thus, a loose Mn, Si, Ni coated Cu/Fe interface is maintained on the Cu side. The ordered phase can then grow by a flux of atoms (that are essentially immediately replaced) from the thin steady-state MnNiSi coating layer, as well as by diffusional fluxes directly to the MnNiSi appendage from the matrix. That is, as additional solutes diffuse from the matrix and the coating layer, prior to solute depletion from the matrix, a steady-state MnNiSi coating layer on the Cu side of the co-precipitate feeding a growing appendage is established.

While this scenario can explain why the appendage structure first forms, it does not explain why the ordered MnNiSi phase does not grow around the Cu core and eliminate the partially coated Cu/Fe interface. Another important observation in the LKMC simulations provides insights on this point. As shown in Figure 4, there exists an absolute movement of the Cu core relative to the center of the whole co-precipitate during the growth process. Given the fact that the lattice sites are conserved in the KLMC simulations, a movement of Cu cluster toward one direction relative to the whole precipitate indicates a net atomic flux in the opposite direction, in this case, composed of Mn, Ni and Si atoms. This observation suggests that during the process of the formation of co-precipitate, apart from the more obvious diffusion path along the Cu/Fe interface, there is another diffusion route for Mn, Ni and Si atoms through the Cu core, i.e., the Cu core does not block the Mn, Ni and Si diffusion but instead serve as diffusion media, like the Fe matrix. This is consistent with the basic vacancy thermodynamics for Fe and Cu. In particular, it has been shown by DFT calculations that compared to in the Fe matrix, the vacancy formation energy in the bcc Cu phase is much lower [36], thus the diffusion of solutes inside the Cu core could be significant. In addition, since the solute atoms would diffuse through the Cu clusters to join the ordered phase on the appendage side of the co-precipitate, the Mn, Ni and Si build-up on the Cu side is limited, suppressing the nucleation of other ordered MnNiSi nuclei, contributing to the stabilization of the appendage structure.

To further confirm the identified mechanisms, we use KLMC simulations to study the dependence of the precipitate structure on two key parameters. First, we note that our model predicts that if the Cu-Fe interface energy is higher, the Mn, Ni, Si coating layer could be more stable, and the energetic penalty of (at least partially) exposing the Cu-Fe interface may dominate over the energy gain by forming the ordered appendage phase. In this case, a core, and ultimately ordered, thick shell may be favored. To test this hypothesis, the KLMC simulation was rerun with an increased Cu-Fe interfacial energy by increasing the Cu-Fe interaction parameter $\Omega_{Cu\text{-}Fe}$ from 0.458eV to 0.758eV; this leads to a > 60% increase in Cu-Fe interfacial energy, according to the NNBB model. The snapshot of the simulations in Figure 5a shows that most of the Cu clusters are fully enveloped by a thick MnNiSi shell. Detailed examination shows that the Cu cores are densely coated by MnNiSi atoms from the beginning of precipitation, essentially eliminating the Fe/Cu interface. A magnified precipitate is shown in Figure 5b, again with a ROI passing through it. A 1-d concentration profile is shown in Figure 5c, showing the core-shell structure. The change in morphology from appendage to core-shell structures that is observed when the interfacial energy contribution is increased supports our hypothesis that the appendage structure occurs due to the larger energy gain by forming an ordered MnNiSi phase compared to the interfacial energy penalty.

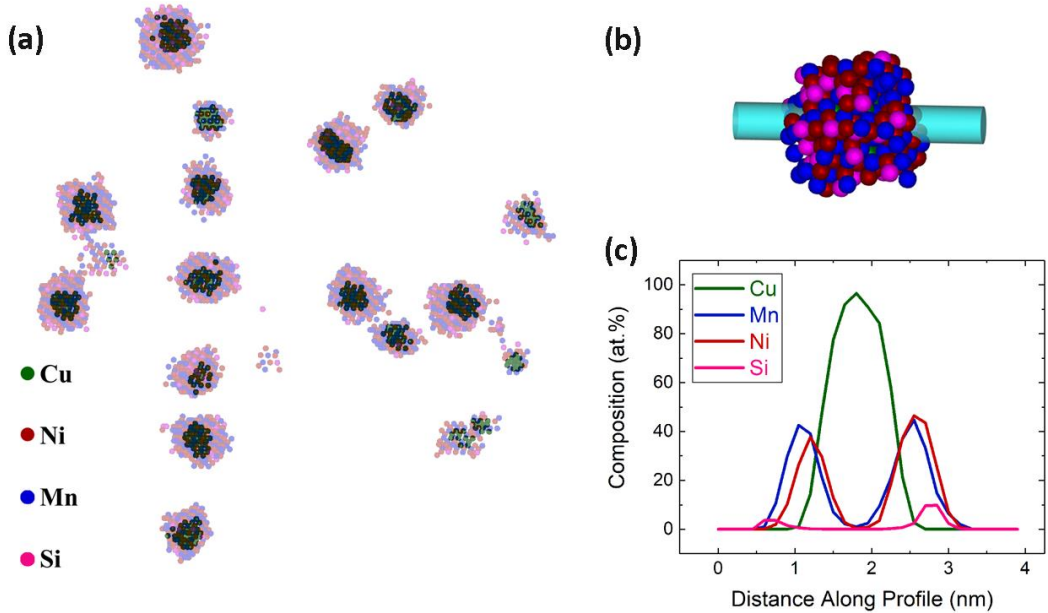

Figure 5 (a) Precipitate morphology for system with an artificially increased Cu-Fe interfacial energy. Cu atoms are shown in solid green; Ni (red), Mn (blue) and Si (magenta) atoms are shown as partially transparent to more clearly see the precipitate morphology. (b) a magnified Cu-MnNiSi precipitate, with a cylindrical ROI along which a 1-D composition profile will be calculated (c) a 1-D composition profile along the z-axis of the ROI. (Color online)

In addition to the test of increasing the Cu-Fe interfacial energy, to confirm the importance of the ordering of MnNiSi phase in the co-precipitate evolution process, we also carried out simulations that artificially suppressed the ordering by randomly swapping the Mn, Ni, and Si atoms. The simulation temperature was lowered to 50˚C to increase the driving force for MnNiSi precipitation. The resulting microstructure is similar to that shown in Figure 5a, manifesting formation of core-shell structure.

Figure 6 summarizes the proposed pathway of the formation of the appendage morphology. In the case of Cu solutes, the large free energy reduction for phase separation and high diffusion rates, resulting fast kinetics, lead to rapid nucleation and initial growth of copper-rich precipitates. The copper-rich precipitates are characterized by a thin disordered Mn-Ni-Si layer at the Cu/Fe boundary that reduces the interfacial energy. Further segregation of Mn, Ni and Si solutes at the Fe/Cu interface then leads to heterogeneous nucleation of the more energetically favorable ordered nucleus of a MnNiSi phase at a facet on the copper rich precipitates, creating the initial asymmetric appendage morphology. Since the ordered phase is energetically favorable compared to the disordered Mn-Ni-Si layer, it grows by interfacial diffusion as well as bulk diffusion through both the Fe matrix and the Cu core. The solute flux through the Cu core is signaled by a net displacement away from the center of the whole co-precipitate during growth, resulting in the appendage morphology observed in both APT reconstructions and KLMC simulations. This observation suggests that during the process of the formation of Cu-MnNiSi precipitate, there is a diffusion path for Mn, Ni and Si atoms through the Cu core. The diffusion paths of the Mn, Ni, Si atoms maintain the asymmetric appendage structure throughout the growth process. Note that there may be an element of dendritic-type growth at the initially small radius tip of the MnNiSi appendage [60], although we believe this mechanisms plays a minimal role for larger precipitates as the combined Cu and MNS precipitate takes an overall spherical shape.

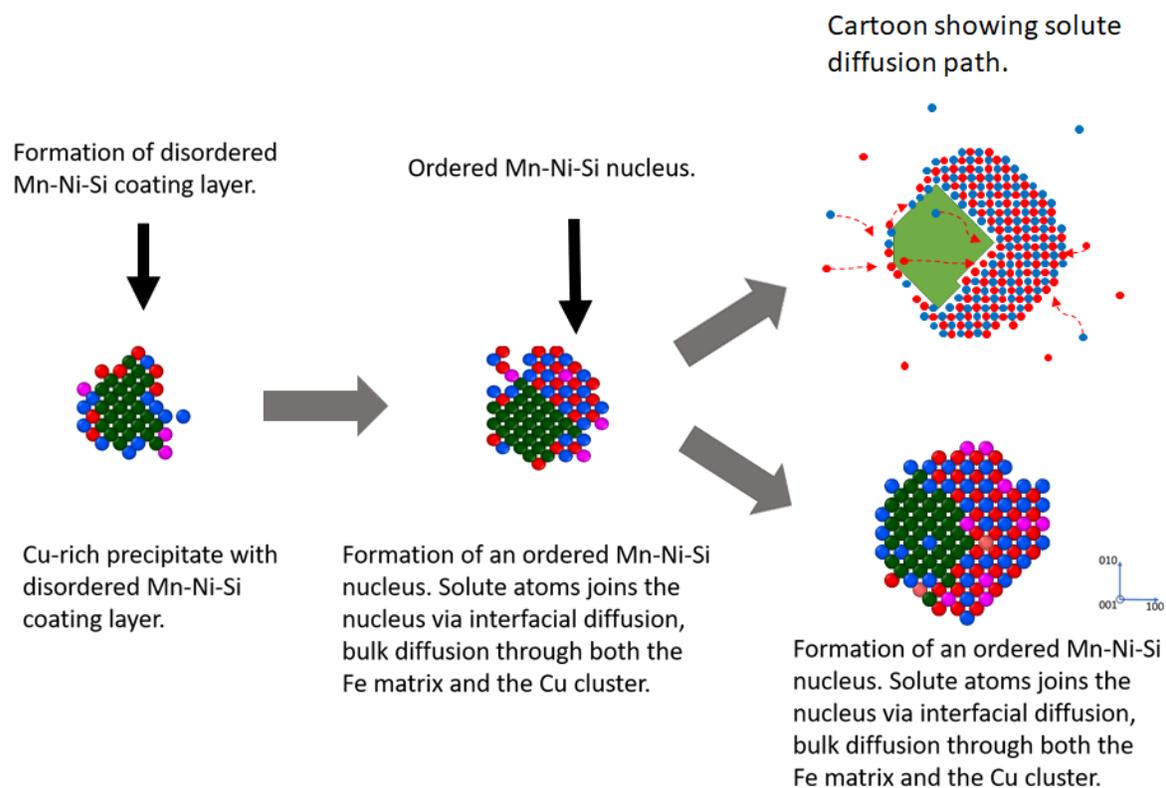

Figure 6. A series of 001 slices of the Cu-MnNiSi precipitate showing the evolution of the co-precipitate. Faceting is observed for the precipitates, consistent with the interfacial energy calculations. Note the specifically illustrated solute diffusion paths in the schematic on the top right corner. For interpretation of the references to color in this figure, the reader is referred to the web version of this article. Green: copper-rich precipitate; blue: Ni atom; red: Mn/Si atom; pink: Fe atom. (color online).

The formation of appendage morphology is not limited to the system discussed in this article. Similar precipitate structures have been observed in other multicomponent alloy systems, such as nanocrystalline FINEMET alloy [4], 12%Cr–9%Ni–4%Mo–2%Cu stainless steel [5], and Cu-bearing low alloy steels [6, 7]. The mechanisms described in this article can be readily applied all examples referred above. In these studies, the system all involve a fast-precipitating, highly immiscible species (namely, Cu) in the matrix, and an ordered second phase ($DO_3$[4], $L1_0$[5], B2[6, 7]) co-precipitating on the fast-forming Cu cluster. Note that, in other systems where the two precipitating phases have the same ordering, such as in Ref. [2] and [61], the slower-precipitating second phase can form, with essentially no nucleation barrier, as an ordered shell on the ordering template provided by the fast-precipitating phase, leading to the common core-shell structure.

## 4. Conclusions

In this study, we use KLMC simulations to elucidate the mechanisms that govern co-precipitate morphologies in multi-constituent alloy systems that have two distinct co-precipitating phases. Here, we simulate an informative example for an FeCuMnNiSi alloy. It is shown that the interplay between interfacial energies, ordering energies, and active diffusion paths determines the co-precipitate morphologies. In this case, the ordering energy of nucleating a discrete associated MnNiSi phase is larger than the energetic penalty of partially exposing the interface of a previously formed Cu core-dilute shell precipitate. The active Mn, Ni, and Si diffusion paths through and around the Cu core enable further steady growth of the MnNiSi appendage. We also demonstrated by KLMC simulations that the favored co-precipitate morphology can be altered from an appendage to a core-shell structure by increasing the interfacial energy between the matrix and the fast-precipitating Cu-rich phase.

The kinetic pathway that we discovered provides both a general guidance in understanding co-precipitation process and insights in designing the co-precipitation microstructures in multi-component alloy systems.

**Acknowledgements**

The US Department of Energy Office of Nuclear Energy's Light Water Reactor Sustainability Program, Materials Aging and Degradation Pathway supported both the University of Wisconsin and USCB contributions to the recent research reported here. The authors gratefully acknowledge use of facilities and instrumentation supported by NSF through the UCSB Materials Research Laboratory Microscopy and Microanalysis Facility (DMR 1121053) for APT measurements.

**References**

[1] Z.B. Jiao, J.H. Luan, M.K. Miller, Y.W. Chung, C.T. Liu, Co-precipitation of nanoscale particles in steels with ultra-high strength for a new era, Materials Today 20(3) (2017) 142-154.
[2] V. Radmilovic, C. Ophus, E.A. Marquis, M.D. Rossell, A. Tolley, A. Gautam, M. Asta, U. Dahmen, Highly monodisperse core-shell particles created by solid-state reactions, Nat. Mater. 10(9) (2011) 710-715.
[3] X. Zhang, N.Q. Vo, P. Bellon, R.S. Averback, Microstructural stability of nanostructured Cu-Nb-W alloys during high-temperature annealing and irradiation, Acta Mater. 59(13) (2011) 5332-5341.
[4] K. Hono, D.H. Ping, M. Ohnuma, H. Onodera, Cu clustering and Si partitioning in the early crystallization stage of an Fe73.5Si13.5B9Nb3Cu1 amorphous alloy, Acta Mater. 47(3) (1999) 997-1006.
[5] M. Hättestrand, J.-O. Nilsson, K. Stiller, P. Liu, M. Andersson, Precipitation hardening in a 12%Cr–9%Ni–4%Mo–2%Cu stainless steel, Acta Mater. 52(4) (2004) 1023-1037.
[6] R. Prakash Kolli, D.N. Seidman, The temporal evolution of the decomposition of a concentrated multicomponent Fe–Cu-based steel, Acta Mater. 56(9) (2008) 2073-2088.
[7] Z.B. Jiao, J.H. Luan, M.K. Miller, C.Y. Yu, C.T. Liu, Group precipitation and age hardening of nanostructured Fe-based alloys with ultra-high strengths, Scientific Reports 6 (2016) 21364.
[8] X. Zhang, S. Shu, P. Bellon, R.S. Averback, Precipitate stability in Cu–Ag–W system under high-temperature irradiation, Acta Mater. 97(0) (2015) 348-356.
[9] H. Leitner, M. Schober, R. Schnitzer, Splitting phenomenon in the precipitation evolution in an Fe–Ni–Al–Ti–Cr stainless steel, Acta Mater. 58(4) (2010) 1261-1269.
[10] S. Shu, X. Zhang, P. Bellon, R.S. Averback, Non-equilibrium Grain Boundary Wetting in Cu–Ag Alloys Containing W Nanoparticles, Materials Research Letters 4(1) (2016) 22-26.
[11] T.F. Kelly, M.K. Miller, Invited review article: Atom probe tomography, Review of Scientific Instruments 78(3) (2007).
[12] D.N. Seidman, Three-dimensional atom-probe tomography: Advances and applications, Annual Review of Materials Research2007, pp. 127-158.

[13] A.F. Voter, INTRODUCTION TO THE KINETIC MONTE CARLO METHOD, in: K.E. Sickafus, E.A. Kotomin, B.P. Uberuaga (Eds.), Radiation Effects in Solids, Springer Netherlands, Dordrecht, 2007, pp. 1-23.
[14] P.B. Wells, T. Yamamoto, B. Miller, T. Milot, J. Cole, Y. Wu, G.R. Odette, Evolution of manganese–nickel–silicon-dominated phases in highly irradiated reactor pressure vessel steels, Acta Mater. 80 (2014) 205-219.
[15] P.D. Styman, J.M. Hyde, K. Wilford, G.D.W. Smith, Quantitative methods for the APT analysis of thermally aged RPV steels, Ultramicroscopy 132 (2013) 258-264.
[16] P.D. Styman, J.M. Hyde, K. Wilford, D. Parfitt, N. Riddle, G.D.W. Smith, Characterisation of interfacial segregation to Cu-enriched precipitates in two thermally aged reactor pressure vessel steel welds, Ultramicroscopy 159, Part 2 (2015) 292-298.
[17] S. Shu, B.D. Wirth, P.B. Wells, D.D. Morgan, G.R. Odette, Multi-technique characterization of the precipitates in thermally aged and neutron irradiated Fe-Cu and Fe-Cu-Mn model alloys: Atom probe tomography reconstruction implications, Acta Mater. 146 (2018) 237-252.
[18] C.L. Liu, G.R. Odette, B.D. Wirth, G.E. Lucas, A lattice Monte Carlo simulation of nanophase compositions and structures in irradiated pressure vessel Fe-Cu-Ni-Mn-Si steels, Materials Science and Engineering: A 238(1) (1997) 202-209.
[19] G.R. Odette, C.L. Liu, B.D. Wirth, On the Composition and Structure of Nanoprecipitates in Irradiated Pressure Vessel Steels, MRS Proceedings 439 (1996).
[20] P.D. Styman, J.M. Hyde, K. Wilford, A. Morley, G.D.W. Smith, Precipitation in long term thermally aged high copper, high nickel model RPV steel welds, Progress in Nuclear Energy 57 (2012) 86-92.
[21] Q. Liu, J. Gu, W. Liu, On the Role of Ni in Cu Precipitation in Multicomponent Steels, Metall and Mat Trans A 44(10) (2013) 4434-4439.
[22] P.D. Edmondson, M.K. Miller, K.A. Powers, R.K. Nanstad, Atom probe tomography characterization of neutron irradiated surveillance samples from the R. E. Ginna reactor pressure vessel, J. Nucl. Mater. 470 (2016) 147-154.
[23] H. Watanabe, S. Arase, T. Yamamoto, P. Wells, T. Onishi, G.R. Odette, Hardening and microstructural evolution of A533b steels irradiated with Fe ions and electrons, J. Nucl. Mater. 471 (2016) 243-250.
[24] K. Lindgren, M. Boåsen, K. Stiller, P. Efsing, M. Thuvander, Evolution of precipitation in reactor pressure vessel steel welds under neutron irradiation, J. Nucl. Mater. 488 (2017) 222-230.
[25] M.K. Miller, Atom probe tomography, Handbook of Microscopy for Nanotechnology (2005) 227-246.
[26] D.J. Larson, T.J. Prosa, R.M. Ulfig, B.P. Geiser, T.F. Kelly, Local electrode atom probe tomography: a user's guide, Springer Science & Business Media2013.
[27] R.A. Enrique, P. Bellon, Compositional patterning in immiscible alloys driven by irradiation, Phys. Rev. B 63(13) (2001).
[28] S. Shu, P. Bellon, R.S. Averback, Role of point-defect sinks on irradiation-induced compositional patterning in model binary alloys, Phys. Rev. B 91(21) (2015) 214107.
[29] G. Martin, P. Bellon, Driven alloys, in: H. Ehrenreich, F. Spaepen (Eds.), Solid State Physics - Advances in Research and Applications, Vol 50, Elsevier Academic Press Inc, San Diego, 1997, pp. 189-331.
[30] S. Shu, P. Bellon, R.S. Averback, Complex nanoprecipitate structures induced by irradiation in immiscible alloy systems, Phys. Rev. B 87(14) (2013).

[31] E. Vincent, C.S. Becquart, C. Pareige, P. Pareige, C. Domain, Precipitation of the FeCu system: A critical review of atomic kinetic Monte Carlo simulations, J. Nucl. Mater. 373(1–3) (2008) 387-401.
[32] L. Messina, M. Nastar, T. Garnier, C. Domain, P. Olsson, Exact ab initio transport coefficients in bcc Fe-X (X=Cr, Cu, Mn, Ni, P, Si) dilute alloys, Phys. Rev. B 90(10) (2014) 104203.
[33] M. Doyama, J.S. Koehler, The relation between the formation energy of a vacancy and the nearest neighbor interactions in pure metals and liquid metals, Acta Metallurgica 24(9) (1976) 871-879.
[34] C. Kittel, Introduction to Solid State Physics, Wiley1995.
[35] E. Vincent, C.S. Becquart, C. Domain, Atomic kinetic Monte Carlo model based on ab initio data: Simulation of microstructural evolution under irradiation of dilute Fe–CuNiMnSi alloys, Nuclear Instruments and Methods in Physics Research Section B: Beam Interactions with Materials and Atoms 255(1) (2007) 78-84.
[36] F. Soisson, C.-C. Fu, Cu-precipitation kinetics in α-Fe from atomistic simulations: Vacancy-trapping effects and Cu-cluster mobility, Phys. Rev. B 76(21) (2007) 214102.
[37] J.O. Andersson, T. Helander, L. Höglund, P. Shi, B. Sundman, Thermo-Calc & DICTRA, computational tools for materials science, Calphad 26(2) (2002) 273-312.
[38] S. Shu, P. Wells, N. Almirall, G.R. Odette, D. Morgan, A kinetic Monte Carlo Study of Post-Irradiation Annealing of Reactor Pressure Vessel Steels.
[39] E. Vincent, C.S. Becquart, C. Domain, Microstructural evolution under high flux irradiation of dilute Fe–CuNiMnSi alloys studied by an atomic kinetic Monte Carlo model accounting for both vacancies and self interstitials, J. Nucl. Mater. 382(2–3) (2008) 154-159.
[40] H. Ullmaier, P. Ehrhart, P. Jung, H. Schultz, Atomic Defects in Metals, Springer Berlin Heidelberg1991.
[41] F. Soisson, T. Jourdan, Radiation-accelerated precipitation in Fe–Cr alloys, Acta Mater. 103 (2016) 870-881.
[42] H. Ke, P. Wells, P.D. Edmondson, N. Almirall, L. Barnard, G.R. Odette, D. Morgan, Thermodynamic and kinetic modeling of Mn-Ni-Si precipitates in low-Cu reactor pressure vessel steels, Acta Mater. 138 (2017) 10-26.
[43] G.R. Odette, T. Yamamoto, D. Klingensmith, On the effect of dose rate on irradiation hardening of RPV steels, Philos. Mag. 85(4-7) (2005) 779-797.
[44] A. Wagner, F. Bergner, R. Chaouadi, H. Hein, M. Hernández-Mayoral, M. Serrano, A. Ulbricht, E. Altstadt, Effect of neutron flux on the characteristics of irradiation-induced nanofeatures and hardening in pressure vessel steels, Acta Mater. 104 (2016) 131-142.
[45] J.M. Hyde, G. Sha, E.A. Marquis, A. Morley, K.B. Wilford, T.J. Williams, A comparison of the structure of solute clusters formed during thermal ageing and irradiation, Ultramicroscopy 111(6) (2011) 664-671.
[46] A. Morley, G. Sha, S. Hirosawa, A. Cerezo, G.D.W. Smith, Determining the composition of small features in atom probe: bcc Cu-rich precipitates in an Fe-rich matrix, Ultramicroscopy 109(5) (2009) 535-540.
[47] P.D. Edmondson, C.M. Parish, R.K. Nanstad, Using complimentary microscopy methods to examine Ni-Mn-Si-precipitates in highly-irradiated reactor pressure vessel steels, Acta Mater. 134 (2017) 31-39.
[48] D. Beinke, C. Oberdorfer, G. Schmitz, Towards an accurate volume reconstruction in atom probe tomography, Ultramicroscopy 165 (2016) 34-41.

[49] F. Vurpillot, C. Oberdorfer, Modeling Atom Probe Tomography: A review, Ultramicroscopy 159 (2015) 202-216.
[50] T. Philippe, M. Gruber, F. Vurpillot, D. Blavette, Clustering and Local Magnification Effects in Atom Probe Tomography: A Statistical Approach, Microsc. microanal. 16(5) (2010) 643-648.
[51] E.A. Marquis, F. Vurpillot, Chromatic Aberrations in the Field Evaporation Behavior of Small Precipitates, Microsc. microanal. 14(06) (2008) 561-570.
[52] W. Xiong, H.B. Ke, R. Krishnamurthy, P. Wells, L. Barnard, G.R. Odette, D. Morgan, Thermodynamic models of low-temperature Mn-Ni-Si precipitation in reactor pressure vessel steels, MRS Communications 4(3) (2014) 101-105.
[53] X. Yan, A. Grytsiv, P. Rogl, V. Pomjakushin, X. Xue, On the crystal structure of the Mn–Ni–Si G-phase, Journal of Alloys and Compounds 469(1) (2009) 152-155.
[54] D.J. Sprouster, J. Sinsheimer, E. Dooryhee, S.K. Ghose, P. Wells, T. Stan, N. Almirall, G.R. Odette, L.E. Ecker, Structural characterization of nanoscale intermetallic precipitates in highly neutron irradiated reactor pressure vessel steels, Scr. Mater. 113 (2016) 18-22.
[55] J. Millán, S. Sandlöbes, A. Al-Zubi, T. Hickel, P. Choi, J. Neugebauer, D. Ponge, D. Raabe, Designing Heusler nanoprecipitates by elastic misfit stabilization in Fe–Mn maraging steels, Acta Mater. 76 (2014) 94-105.
[56] C. Zhang, M. Enomoto, Study of the influence of alloying elements on Cu precipitation in steel by non-classical nucleation theory, Acta Mater. 54(16) (2006) 4183-4191.
[57] P. Wells, H.B. Ke, N. Almirall, D. Morgan, T. Yamamoto, G.R. Odette, On the Thermal Stability of Features Formed in Highly Irradiated Reactor Pressure Vessel Steels, (in preparation).
[58] B. Sonderegger, E. Kozeschnik, Generalized Nearest-Neighbor Broken-Bond Analysis of Randomly Oriented Coherent Interfaces in Multicomponent Fcc and Bcc Structures, Metallurgical and Materials Transactions a-Physical Metallurgy and Materials Science 40A(3) (2009) 499-510.
[59] O. Pierre-Louis, Solid-state wetting at the nanoscale, Progress in Crystal Growth and Characterization of Materials 62(2) (2016) 177-202.
[60] D.A. Porter, K.E. Easterling, M. Sherif, Phase Transformations in Metals and Alloys, (Revised Reprint), CRC press2009.
[61] E. Clouet, L. Lae, T. Epicier, W. Lefebvre, M. Nastar, A. Deschamps, Complex precipitation pathways in multicomponent alloys, Nat. Mater. 5(6) (2006) 482-488.